\documentclass[conference,10pt]{IEEEtran}
\pdfoutput=1
\usepackage[boxruled]{algorithm2e}
\usepackage{authblk}
\usepackage{cite}
\usepackage{graphicx}
\usepackage{psfrag}
\usepackage{subfigure}
\usepackage{url}
\usepackage{amsmath}
\usepackage{array}
\usepackage{amssymb}
\usepackage{amsfonts}
\usepackage{graphicx}
\usepackage{epstopdf}
\usepackage{placeins}
\usepackage{algorithmic}

\newtheorem{lemma}{Lemma}

\newtheorem{definition}{Definition}
\newtheorem{theorem}{Theorem}

\newtheorem{example}{Example}

\usepackage{setspace}
\usepackage{amscd}
\usepackage{multirow}
\usepackage{mathrsfs}
\usepackage{epsfig}
\usepackage{color}
\usepackage{textcomp}
\usepackage{multirow}

\input{epsf.sty}

\title{Vector Linear Error Correcting Index Codes and Discrete Polymatroids}
\begin{document}
\author{Anoop Thomas and B. Sundar Rajan
}
\affil{Dept. of ECE, IISc, Bangalore 560012, India, Email: $\lbrace$anoopt,bsrajan$\rbrace$@ece.iisc.ernet.in.}
\maketitle
\thispagestyle{empty}	
\begin{abstract}
The connection between index coding and matroid theory have been well studied in the recent past. El Rouayheb \textit{et al.} established a connection between multi linear representation of matroids and wireless index coding. Muralidharan and Rajan showed that a vector linear solution to an index coding problem exists if and only if there exists a representable discrete polymatroid satisfying certain conditions. Recently index coding with erroneous transmission was considered by Dau \textit{et al.}. Error correcting index codes in which all receivers are able to correct a fixed number of errors was studied. In this paper we consider a more general scenario in which each receiver is able to correct a desired number of errors, calling such index codes  \textit{differential error correcting index codes.} We show that vector linear differential error correcting index code exists if and only if there exists a representable discrete polymatroid satisfying certain conditions.
\end{abstract}

\section{Introduction}
\label{Sec:Introduction}

The index coding problem introduced by Birk and Kol \cite{birk1998informed} involves a source which generates a set of messages and set of receivers which demand messages. Each receiver has prior knowledge of a portion of the message called side-information. The source uses the side-information available at all the receivers to find a transmission scheme of minimum number of transmissions, which satisfies all the demands of the receivers. Bar-Yossef \textit{et al.} \cite{bar2011index} studied the index coding problem and found that the length of the optimal linear index code is equal to the minrank of a related graph. Lubetzky and Stav \cite{lubetzky2009nonlinear} showed that non-linear scalar codes are better than linear scalar ones. The connection between multi-linear representation of matroids and index coding was studied by El Rouayheb, Sprinston and Georghiades \cite{el2010index}. It was shown in \cite{muralidharan2014linear} that a vector linear solution to an index coding problem exists if and only if there exists a representable discrete polymatroid satisfying certain conditions which are determined by the index coding problem.

The problem of index coding with erroneous transmissions was studied by Dau \textit{et al.} \cite{dau2013error}. An index code capable of correcting at most $\delta$-errors at all its receivers is defined as a $\delta$-error correcting index code. The necessary and sufficient conditions for a scalar linear index code to have $\delta$-error correcting capability was found. Linear network error-correcting codes were introduced earlier by Yeung and Cai \cite{yeung2006network},\cite{cai2006network}. The link between network error correcting codes and certain matroids was established by Prasad and Rajan in \cite{DBLP:conf/isit/PrasadR12}.

 In this paper we consider differential error correcting index codes which allows receivers to have different error correcting capability. We establish a link between vector linear differential error correcting index codes and discrete polymatroids. We show that a vector linear solution to an error correcting index coding problem exists if and only if there exists a representable discrete polymatroid satisfying certain conditions which are determined by the index coding problem considered. Error correction at a subset of receivers and $\delta$-error correcting index codes are also considered and the representable discrete polymatroids associated with these cases are identified. 

The organization of the paper is as follows. In Section \ref{Sec:ErrorCorrectingIndexCodes} we review error correcting index codes and also establish a lemma which is used to prove our main result. In Section \ref{Sec:DiscretePolymatroid}, basic results of discrete polymatroids are reviewed. Finally in Section \ref{Sec:ECICandDPM}, we establish the connection between vector error correcting index codes and discrete polymatroids. We conclude and summarize our results in Section \ref{Sec:Conclusion}.

\textbf{\textit{Notations:}}
The set $\lbrace 1,2,\dotso,m \rbrace$ is denoted as  $\lceil m \rfloor.$ For two sets $S_{1}$ and $S_{2}$ the set subtraction $S_{1} \setminus S_{2}$ is denoted by $S_{1}-S_{2}$. $\mathbb{Z}_{\geq 0} $ denotes the set of non-negative integer. For a positive integer $n$, $\underline{\mathbf{0}}_{n}$ denotes all zero vector of length $n$. For a vector $v$ of length $m$ and $A \subseteq \lceil m \rfloor,$ $v_{A}$ is the vector obtained by taking only the components of $v$ indexed by the elements of $A.$ The vector of length $m$ whose $i^{\text{th}}$ component is one and all other components are zeros is denoted as $\epsilon_{i,m}.$  For $u,v \in \mathbb{Z}_{\geq 0}^m,$ $u \leq v$ if all the components of $v-u$ are non-negative and,  $u < v$ if $u \leq v$ and $u \neq v.$ For $u,v \in \mathbb{Z}_{\geq 0}^m,$ $u \vee v$ is the vector whose $i^{\text{th}}$ component is the maximum of the $i^{\text{th}}$ components of $u$ and $v.$ A vector $u \in \mathbb{Z}_{\geq 0}^m$ is called an integral sub-vector of $v \in \mathbb{RZ}_{\geq 0}^m$ if $u \leq v.$ For a vector $u \in \mathbb{Z}_{\geq 0}^{m},$ $(u)_{>0}$ denotes the set of indices corresponding to the non-zero components of $u.$ The magnitude of a vector $v \in  \mathbb{Z}_{\geq 0}^r,$ is the sum of the components of $v$ and is denoted by $\vert v \vert$. The \textit{support} of vector $x \in \mathbb{F}_{q}^{n}$ is defined to be the set $supp(x)=\lbrace i \in \lceil n \rfloor : x_{i} \neq 0 \rbrace.$ The Hamming weight of a vector $x$, denoted by $wt(x)$, is defined to be the $|supp(x)|$. The rank of a matrix $A$ over $\mathbb{F}_{q}$ is denoted by $rank(A)$. For some positive integer $c$, identity matrix of size $c \times c$ over $\mathbb{F}_{q}$ is denoted by $I_{c}$. The vector space spanned by columns of a matrix $A$ over $\mathbb{F}_{q}$ is denoted by $\langle A \rangle$. For some matrix $A$, $A^{(i)}$ denotes the $i^{th}$ column of $A$. For a set of column indices $\mathcal{I}$, $A^{\mathcal{I}}$ denotes the submatrix of $A$ with columns indexed by $\mathcal{I}$. Similarly $A_{(j)}$ denotes the $j^{th}$ row of $A$ and for a set of row indices $\mathcal{J}$, $A_{\mathcal{J}}$ denotes the submatrix of $A$ with rows indexed by $\mathcal{J}$.

\section{Error Correcting Index Codes and a useful Lemma}
\label{Sec:ErrorCorrectingIndexCodes}

An index coding problem $\mathcal{I}(X,\mathcal{R})$ includes
\begin{itemize}
\item
a set of messages $X=\{x_1,x_2,\dotso,x_m\}$ and 
\item
a set of receiver nodes $\mathcal{R}\subseteq \{(x,H);x \in X, H \subseteq X \setminus \{x\}\}.$
\end{itemize}
For a receiver node $R_{i}=(x_{f(i)},H_{i})\in \mathcal{R},$ $x_{f(i)}$ denotes the message demanded by receiver $R_{i}$ and $H_{i} \subseteq X -\lbrace x_{f(i)} \rbrace$ denotes the side information possessed by $R_{i}.$ Note that $f$ is a mapping from $\left\lceil \vert \mathcal{R} \vert \right\rfloor $ to $ \left\lceil m \right\rfloor$. Each one of the messages $x_i, i \in \{1,2,\dotso,m\},$ are assumed to be row vectors of length $n,$ over an alphabet set, which in this paper is assumed to be a finite field $\mathbb{F}_q$ of size $q.$   Let $y=[x_1\;x_2\dotso x_m]$ denote the row vector of length $nm$ obtained by the concatenation of the $m$ message vectors.

An index coding solution (also referred to as an index code) over $\mathbb{F}_q$ of length $c$ and dimension $n$ for the index coding problem $\mathcal{I}(X,\mathcal{R})$ is a function $\mathfrak{C} : \mathbb{F}_q^{mn} \rightarrow \mathbb{F}_q^{c},$ $c$ an integer, which satisfies the following condition: For every $R_{i}=(x_{f(i)},H_{i}) \in \mathcal{R},$ there exists a function $\psi_{R_{i}}: F_q^{n \vert H_{i} \vert +c} \rightarrow F_q^{n}$ such that $\displaystyle{\psi_{R_{i}}(H_{i},\mathfrak{C}(y))=x_{f(i)},\forall y \in \mathbb{F}^{mn}_q.}$
The function $\psi_{R_{i}}$ is referred to as the decoding function at receiver $R_{i}.$

An index coding solution for which $n=1$ is called a scalar solution; otherwise it is called a vector solution. An index coding solution is said to be linear if the encoding and decoding functions are linear. When the index coding solution is linear it can be described as  $ \mathfrak{C}(y)=yL,$  $\forall ~y \in \mathbb{F}_{q}^{mn},$ where $L$ is a $mn\times c$ matrix over $\mathbb{F}_{q}$. The matrix $L$ is called the matrix corresponding to the linear index code $\mathfrak{C}$. The code $\mathfrak{C}$ is referred to as the linear index code based on $L$.

Error correcting index code considers the scenario in which the symbols received by receivers may be subject to errors. The source $S$ broadcasts a vector $\mathfrak{C}(y) \in \mathbb{F}_{q}^{c}$. Consider a receiver $R_{i}=(x_{f(i)},H_{i}) \in \mathcal{R}$. The error affecting receiver $R_{i}$ is considered as an additive error represented by $\epsilon_{i} \in \mathbb{F}_{q}^{c}$. Then, receiver $R_{i}$ actually receives the vector \[y'_{i}=\mathfrak{C}(y)+\epsilon_{i} \in \mathbb{F}_{q}^{c}.\] An error correcting index code should be able to satisfy the demands of all the receivers in the presence of these additive errors. Consider an instance of the index coding problem described by $\mathcal{I}(X,\mathcal{R})$. A \textit{$\delta$-error correcting index code} ($\delta$-ECIC) over $\mathbb{F}_{q}$ for this instance is an encoding function \[ \mathfrak{C}:\mathbb{F}_{q}^{mn} \rightarrow \mathbb{F}_{q}^c \] such that for every receiver $R_{i} = (x_{f(i)},H_{i}) \in \mathcal{R}$, there exists a decoding function $\psi_{R_{i}}: F_q^{n \vert H \vert +c} \rightarrow F_q^{n}$ satisfying \[ \psi_{R_{i}}(\mathfrak{C}(y)+ \epsilon_{i},H_{i})=x_{f(i)}, 
 ~\forall ~ y \in \mathbb{F}_{q}^{mn}, \forall ~ \epsilon_{i} \in \mathbb{F}_{q}^{c}, wt(\epsilon_{i})\leq \delta. \]

Similar to the index coding solution, if the functions $\mathfrak{C}$ and $\psi_{R_i}$ are linear then it is said to be a linear $\delta$-error correcting index code. A linear error correcting index code can also be described by a matrix. Dau \textit{et al.} in \cite{dau2013error} identify a necessary and sufficient condition which a matrix $L$ has to satisfy to correspond to a $\delta$-error correcting index code. However the index coding solution considered  in that paper is a scalar solution. It was observed in \cite{dau2012security} that if the block length is fixed one can model a vector index code as a scalar index code applied to another instance of the index coding problem. If the block length is $n$, the number of messages is $m$, and the number of receivers is $|\mathcal{R}|$ in the vector index coding problem, then the equivalent scalar index coding problem will have $mn$ messages and $n|\mathcal{R}|$ receivers. Using this observation a necessary and sufficient conditions which matrix $L$ has to satisfy to correspond to a vector $\delta$-error correcting index code can be found. We consider a more general error correcting index coding problem in which the receivers have different error correcting capability. Each receiver $R_{i}=(x_{f(i)},H_{i}) \in \mathcal{R}$ should be able to correct $\delta_{i}$ errors. Such index codes are referred to as differential error correcting index codes. 

Consider an instance of the index coding problem described by $\mathcal{I}(X,\mathcal{R})$. Let $\delta_{i}$ be the maximum number of errors receiver $R_{i}$ wants to correct. A \textit{differential error correcting index code} over $\mathbb{F}_{q}$ for this instance is an encoding function \[ \mathfrak{C}:\mathbb{F}_{q}^{mn} \rightarrow \mathbb{F}_{q}^c \] such that for every receiver $R_{i} = (x_{f(i)},H_{i}) \in \mathcal{R}$, there exists a decoding function $\psi_{R_{i}}: F_q^{n \vert H \vert +c} \rightarrow F_q^{n}$ satisfying \[ \psi_{R_{i}}(\mathfrak{C}(y)+ \epsilon_{i},H_{i})=x_{f(i)}, 
 ~\forall ~ y \in \mathbb{F}_{q}^{mn}, \forall ~ \epsilon_{i} \in \mathbb{F}_{q}^{c}, wt(\epsilon_{i})\leq \delta_{i}. \]

A linear differential error correcting index code can also be described by a matrix. We identify the necessary and sufficient conditions which a matrix $L$ has to satisfy to correspond to a differential error correcting index code. Consider a receiver $R_{i}=(x_{f(i)},H_{i}) \in \mathcal{R}$ of the index coding problem $\mathcal{I}(X,\mathcal{R})$. Let $$\hat{H_{i}}=\underset{k:x_{k}\in H_{i}}{\cup} \lbrace (k-1)n+1,(k-1)n+2,\ldots,kn \rbrace,$$ $$\hat{x_{f(i)}}=\lbrace (f(i)-1)n+1,(f(i)-1)n+2,\ldots,f(i)n \rbrace.$$  Let $\overline{H_{i}}=\lceil m \rfloor - \lbrace j : x_{j} \in  H_{i} \rbrace$ and $\overline{\hat{H_{i}}}$ denote the set $\lceil mn \rfloor - \hat{H_{i}}$. A matrix $L$ corresponds to a differential error correcting index code if and only if the following condition is satisfied : for every receiver $R_{i}=(x_{f(i)},H_{i}) \in \mathcal{R}$ and for all $y \in \mathbb{F}_{q}^{mn}$ such that $y_{\hat{x_{f(i)}}} \neq \underline{\mathbf{0}}_{\vert \hat{x_{f(i)}} \vert}$ and $y_{\hat{H_{i}}}=\underline{\mathbf{0}}_{\vert \hat{H_{i}} \vert}$,\begin{equation}
\label{eq:EC1}
yL+\epsilon \neq 0, ~~~ \forall \epsilon \in \mathbb{F}_{q}^{c}, wt(\epsilon) \leq 2\delta_{i}.
\end{equation} 
In the rest of the paper, the subscripts under the zero vector is removed, and the appropriate size of the zero vector is understood from the context. This equation can be rewritten in matrix form in the following way. For each receiver $R_{i}$,
\begin{flalign}
\begin{split}
\label{eq:ECMatrixRepresentation}
(y\quad \epsilon)\left(\begin{array}{c}
L\\
I_{c}
\end{array}\right)\neq 0, 
\end{split}
\end{flalign} 
for all $ ~ y \in \mathbb{F}_{q}^{mn}$ such that $y_{\hat{H_{i}}}=\underline{\mathbf{0}},y_{\hat{x_{f(i)}}} \neq \underline{\mathbf{0}},$ and for all $ ~ \epsilon \in \mathbb{F}_{q}^{c}$ such that $ wt(\epsilon)\leq 2\delta_{i}.$

The \textit{error pattern} corresponding to an error vector $\epsilon$ is defined as its support set $supp(\epsilon)$. Let $\mathbb{I}_{supp(\epsilon)}$ denote the submatrix of $I_{c}$ consisting of those rows of $I_{c}$ indexed by $supp(\epsilon)$. For a receiver $R_{i}$, the error correcting condition \eqref{eq:ECMatrixRepresentation} can be rewritten as 
\begin{equation}
\begin{split}
\label{eq:EC2}	 	
(y \quad \overline{\epsilon})\left(\begin{array}{c}
L\\
\mathbb{I}_{supp(\epsilon)} 
\end{array}\right)\neq 0, \forall ~ y \in \mathbb{F}_{q}^{mn}: y_{\hat{H_{i}}} = \underline{\mathbf{0}},y_{\hat{x_{f(i)}}} \neq \underline{\mathbf{0}}, \\ \forall ~ \overline{\epsilon} \in \mathbb{F}_{q}^{2\delta_{i}}, \forall ~ supp(\epsilon) \in \lbrace \mathcal{F}\subseteq \lceil c \rfloor : |\mathcal{F}|=2\delta_{i} \rbrace.
\end{split}
\end{equation}
So the matrix $L$ corresponds to a $\delta$-error correcting index code if and only if \eqref{eq:EC2} is satisfied at all receivers. Since at a particular receiver we consider only those $y \in \mathbb{F}_{q}^{mn}$ for which $y_{\hat{H_{i}}}=\underline{\mathbf{0}}$, condition \eqref{eq:EC2} can be rewritten as 
\begin{flalign}
\begin{split}
\label{eq:EC3}
(y_{\overline{\hat{H_{i}}}}\quad \overline{\epsilon}) & \left(\begin{array}{c}
L_{\overline{\hat{H_{i}}}}\\
\mathbb{I}_{supp(\epsilon)}
\end{array}\right)\neq 0, \forall ~ y_{\overline{\hat{H_{i}}}} \in \mathbb{F}_{q}^{|\overline{\hat{H_{i}}}|} : y_{\hat{x_{f(i)}}} \neq \underline{\mathbf{0}}, \\ & \forall ~ \overline{\epsilon} \in \mathbb{F}_{q}^{2\delta_{i}}, \forall ~ supp(\epsilon) \in \lbrace \mathcal{F}\subseteq \lceil c \rfloor : |\mathcal{F}|=2\delta_{i} \rbrace.
\end{split}
\end{flalign}

We now present a lemma which will be used in Section \ref{Sec:ECICandDPM} to prove the main result of this paper.

\begin{lemma}
\label{lemma:errorcorrection}
Let $\mathcal{I}_{D(R_{i})}$ denote the  $(|\overline{\hat{H_{i}}}|+2\delta_{i}) \times n$ matrix with the $n \times n$ identity submatrix in $n$ rows of the first $|\overline{\hat{H_{i}}}|$ rows corresponding to the demand $x_{f(i)}$ of the receiver $R_{i}=(x_{f(i)},H_{i})$, with all other elements being zero. For some $ supp(\epsilon) \in \lbrace \mathcal{F}\subseteq \lceil c \rfloor : |\mathcal{F}|=2\delta_{i} \rbrace$ the condition 
\begin{equation}
\begin{split}
(y \quad \overline{\epsilon})\left(\begin{array}{c}
L\\
\mathbb{I}_{supp(\epsilon)} 
\end{array}\right)\neq 0, \forall ~ y \in \mathbb{F}_{q}^{mn}: y_{\hat{H_{i}}} = \underline{\mathbf{0}},y_{\hat{x_{f(i)}}} \neq \underline{\mathbf{0}}, \\ \forall ~ \overline{\epsilon} \in \mathbb{F}_{q}^{2\delta_{i}}, \forall ~ supp(\epsilon) \in \lbrace \mathcal{F}\subseteq \lceil c \rfloor : |\mathcal{F}|=2\delta_{i} \rbrace.
\end{split}
\end{equation}
holds if and only if the following condition holds 
\begin{equation}
\label{eq:EC4}
\mathcal{I}_{D(R_{i})}^{(k)}\subseteq \left\langle \left(\begin{array}{c}
L_{\overline{\hat{H_{i}}}}\\
\mathbb{I}_{supp(\epsilon)}
\end{array}\right)\right\rangle, ~ \forall ~ k \in \lceil n \rfloor. 
\end{equation}
\begin{IEEEproof}
The \textit{if} part is proved first. Since the columns of the matrix $\mathcal{I}_{D(R_{i})}$ is in the subspace  $\left\langle \left(\begin{array}{c}
L_{\overline{\hat{H_{i}}}}\\
\mathbb{I}_{supp(\epsilon)}
\end{array}\right)\right\rangle$, linear combinations of columns of $ \left(\begin{array}{c}
L_{\overline{\hat{H_{i}}}}\\
\mathbb{I}_{supp(\epsilon)}
\end{array}\right)$ should generate $\mathcal{I}_{D(R_{i})}$. There should be some $c \times n$ matrix $X$ such that, \[\left(\begin{array}{c}
L_{\overline{\hat{H_{i}}}}\\
\mathbb{I}_{supp(\epsilon)}
\end{array}\right) X= \mathcal{I}_{D(R_{i})}. \] Now suppose for some $(y \quad \overline{\epsilon})$, with $y_{\hat{x_{f(i)}}}\neq \underline{\mathbf{0}},y_{\hat{H_{i}}}=\underline{\mathbf{0}}$ and some $\overline{\epsilon} \in \mathbb{F}_{q}^{2\delta}$ we have \[
(y \quad \overline{\epsilon})\left(\begin{array}{c}
L\\
\mathbb{I}_{supp(\epsilon)} 
\end{array}\right) = 0. \] Since $y_{\hat{H_{i}}}=\underline{\mathbf{0}}$, the above equation reduces to \[
(y_{\overline{\hat{H_{i}}}}\quad \overline{\epsilon}) \left(\begin{array}{c}
L_{\overline{\hat{H_{i}}}}\\
\mathbb{I}_{supp(\epsilon)}
\end{array}\right)= 0. \] Multiplying both sides by $X$, we get $y_{\hat{x_{f(i)}}}=\underline{\mathbf{0}}$, which is a contradiction. This completes the if part. \\
Now we prove the ``only if''  part. Let $L_{\hat{x_{f(i)}}}$ denote the rows of  $L$ corresponding to the message demanded by receiver $R_{i}$. Let $\overline{r_{i}}$ denote the set $\lceil mn \rfloor - \hat{H_{i}} - \hat{x_{f(i)}}$. Let $L_{\overline{r_{i}}} $ denote the submatrix of $L$ with rows indexed by the set $\overline{r_{i}}$. Because (\ref{eq:EC3}) holds, we have \begin{eqnarray*}
rank\left(\begin{array}{c}
L_{\overline{\hat{H_{i}}}}\\
\mathbb{I}_{supp(\epsilon)}
\end{array}\right) & = & rank\left(\begin{array}{c}
L_{\hat{x_{f(i)}}}\\
L_{\overline{r_{i}}}\\
\mathbb{I}_{supp(e)}
\end{array}\right)\\
 & = & rank\left(L_{\hat{x_{f(i)}}}\right)+rank\left(\begin{array}{c}
L_{\overline{r_{i}}}\\
\mathbb{I}_{supp(e)}
\end{array}\right)\\
 & = & n+rank\left(\begin{array}{c}
L_{\overline{r_{i}}}\\
\mathbb{I}_{supp(e)}
\end{array}\right). 
\end{eqnarray*}
Consider the concatenated matrix $\left(\begin{array}{cc}
L_{\overline{\hat{H_{i}}}}\\
 & I_{D(R_{i})}\\
\mathbb{I}_{supp(e)}
\end{array}\right)$, denoted by $Y$. We have,
\begin{eqnarray*}
rank\left(Y\right) & = & rank\left(\begin{array}{cc}
L_{\hat{x_{f(i)}}} & I_{n}\end{array}\right)+rank\left(\begin{array}{c}
L_{\overline{r_{i}}}\\
\mathbb{I}_{supp(e)}
\end{array}\right)\\
 & = & n+rank\left(\begin{array}{c}
L_{\overline{r_{i}}}\\
\mathbb{I}_{supp(e)}
\end{array}\right).
\end{eqnarray*}
 The concatenated matrix $Y$ has the same rank as the matrix  $\left(\begin{array}{c}
L_{\overline{H_{i}}}\\
\mathbb{I}_{supp(e)}
\end{array}\right)$. This proves the ``only if'' part. 
\end{IEEEproof}
\end{lemma}

Lemma \ref{lemma:errorcorrection} gives an equivalent condition for equation \eqref{eq:EC2}. Therefore a given index code is differential error correcting if and only if \eqref{eq:EC4} holds for all $ supp(e) \in \lbrace \mathcal{F}\subseteq \lceil c \rfloor : |\mathcal{F}|=2\delta_{i} \rbrace$ and at all receivers $R_{i} \in \mathcal{R}$.

\section{Discrete Polymatroid}
\label{Sec:DiscretePolymatroid}

In this section we review the definitions and results from discrete polymatroids. A discrete polymatroid $\mathbb{D}$ is defined as follows:
\begin{definition}[\cite{herzog2002discrete}]
A discrete polymatroid $\mathbb{D}$ on the ground set $\lceil m \rfloor$ is a non-empty finite set of vectors in $\mathbb{Z}_{\geq 0}^m$ satisfying the following conditions:
\begin{itemize}
\item
If $u \in \mathbb{D}$ and $v <u,$ then $v \in \mathbb{D}.$ 
\item
 For all $u, v \in \mathbb{D}$ with $\vert u\vert < \vert v\vert,$
there exists $w \in  \mathbb{D}$ such that $u < w  \leq  u \vee v.$
\end{itemize}
\end{definition}

Let $2^{\lceil m \rfloor}$ denote the power set of the set $\lceil m \rfloor$. For a discrete polymatroid $\mathbb{D},$ the rank function $r^{\mathbb{D}}: 2^{\lceil m \rfloor} \rightarrow \mathbb{Z}_{\geq 0}$ is defined as $r^{\mathbb{D}}(A)=\max \{ \vert u(A) \vert , u \in \mathbb{D}\},$ where $\emptyset \neq A  \subseteq \lceil m \rfloor$ and $r^{\mathbb{D}}(\emptyset)=0.$ Alternatively, a discrete polymatroid $\mathbb{D}$ can be written in terms of its rank function as $\mathbb{D}=\lbrace x \in \mathbb{Z}_{\geq 0}^m: \vert x(A) \vert \leq r^{\mathbb{D}}(A), \forall A \subseteq \lceil m \rfloor \rbrace.$ In the rest of the paper, the superscript $\mathbb{D}$ in $r^{\mathbb{D}}$ is dropped. A discrete polymatroid is completely described by the rank function. So discrete polymatroid $\mathbb{D}$ on $\lceil m \rfloor$ is also denoted by $(\lceil m \rfloor, r)$. The ground set of discrete polymatroid is also denoted by $E(\mathbb{D})$. 

A function $r: 2^{\lceil m \rfloor} \rightarrow \mathbb{Z}_{\geq 0}$ is the rank function of a discrete polymatroid iff it satisfies the following conditions \cite{farras2007ideal}:
\begin{description}
\item [(D1)]
For $A \subseteq B \subseteq \lceil m \rfloor,$ $r(A)\leq r(B).$
\item [(D2)]
 $\forall A,B \subseteq \lceil m \rfloor,$  $r(A \cup B) + r(A\cap B)\leq r(A)+r(B).$
\item [(D3)]
$r(\emptyset)=0.$
\end{description}

A vector $u \in  \mathbb{D}$ for which there does not exist $v \in \mathbb{D}$ such that $u<v,$ is called a basis vector of $\mathbb{D}.$  Let $\mathcal{B}(\mathbb{D})$ denote the set of basis vectors of $\mathbb{D}.$  The sum of the components of a basis vector of $\mathbb{D}$ is referred to as the rank of $\mathbb{D},$ denoted by $rank(\mathbb{D}).$ Note that for all the basis vectors, sum of the components will be equal \cite{vladoiu2006discrete}. A discrete polymatroid is nothing but the set of all integral subvectors of its basis vectors.

\begin{example}
\label{eg:DPM1}
Consider the discrete polymatroid $\mathbb{D}$ on the ground set $\lceil 3 \rfloor$ with rank function $r$ given by $r(\lbrace 1 \rbrace)=r(\lbrace 2 \rbrace)=r(\lbrace 2,3 \rbrace)=2,r(\lbrace 3 \rbrace)=1$ and $r(\lbrace 1,2 \rbrace)=r(\lbrace 1,3 \rbrace)=r(\lbrace 1,2,3 \rbrace)=3.$ The set of basis vectors for this discrete polymatroid is given by $\mathcal{B}(\mathbb{D})=\lbrace (1,1,1),(1,2,0),(2,0,1),(2,1,0) \rbrace$.
\end{example}

\begin{definition}[\cite{farras2007ideal}]
A discrete polymatroid $\mathbb{D}$ is said to be representable over $\mathbb{F}_q$ if there exists vector subspaces $V_1,V_2,\dotso,V_m$ of a vector space $E$ over $\mathbb{F}_q$ such that $dim(\sum_{i \in X} V_i)=r(X),$ $\forall X \subseteq \lceil m \rfloor.$ The set of vector subspaces $V_i,i\in\lceil m \rfloor,$ is said to form a representation of $\mathbb{D}.$ 
A discrete polymatroid is said to be representable if it is representable over some field.
\end{definition}

$\mathbb{D}(V_{1},V_{2}, \ldots, V_{m})$ denotes a representable discrete polymatroid on $\lceil m \rfloor$ with $V_{1},V_{2},\ldots,V_{m}$ as its representation.  Each vector space $V_{i}$ can be described by a matrix $A_{i}$ whose columns span $V_{i}$.

\begin{example}
\label{eg:RDPM1}
Let $A_{1}=\left[\begin{array}{cc}
1 & 0\\
0 & 1\\
0 & 0
\end{array}\right], A_{2}=\left[\begin{array}{cc}
0 & 1\\
0 & 1\\
1 & 1
\end{array}\right]$ and $A_{3}= \left[\begin{array}{cc}
0 \\
0 \\
1 
\end{array}\right]$ be matrices over $\mathbb{F}_{2}$. Let $V_{i}, i \in \lceil 3 \rfloor$ denote the column span of $A_{i}$. The vector subspaces $V_{i}, i \in \lceil 3 \rfloor$ forms a representation over $\mathbb{F}_{2}$ of the discrete polymatroid in Example \ref{eg:DPM1}.
\end{example}

\begin{example}
\label{eg:DPM2}
Let $A_{1}= \left[\begin{array}{cc}
1 \\
0 \\
0 
\end{array}\right],A_{2}= \left[\begin{array}{cc}
0 \\
1 \\
0 
\end{array}\right],A_{3}= \left[\begin{array}{cc}
0 \\
0 \\
1 
\end{array}\right]$ and $A_{4}=\left[\begin{array}{cc}
1 & 0\\
0 & 1\\
0 & 1
\end{array}\right]$ be matrices over $\mathbb{F}_{q}$. Let $V_{i}$ denote the column span of $A_{i}, i \in \lceil 4 \rfloor$. The rank function $r$ of the discrete polymatroid $\mathbb{D}(V_{1},V_{2},V_{3},V_{4})$ is as follows:  $r(X)=1,$ if $X \in \lbrace \lbrace 1 \rbrace,\lbrace 2 \rbrace,\lbrace 3 \rbrace \rbrace$; $r(X)=2$ if $X \in \lbrace \lbrace 1,2 \rbrace,\lbrace 1,3 \rbrace,\lbrace 1,4 \rbrace,\lbrace 2,3 \rbrace,\lbrace 4 \rbrace$ and  $r(X)=3$ otherwise. The set of basis vectors for this discrete polymatroid is given by, 
\begin{flalign*}
\lbrace (0,0,1,2),(0,1,0,2),(0,1,1,1),(1,0,1,1),(1,1,0,1), \\(1,1,1,0)\rbrace.
\end{flalign*}
\end{example}

\begin{example}
Let $r : 2^{\lceil 4 \rfloor} \rightarrow \mathbb{Z}_{\geq 0}$ be a function given by $r(\lbrace 1 \rbrace)=r(\lbrace 2 \rbrace)=r(\lbrace 3 \rbrace)=r(\lbrace 4 \rbrace)=2,r(\lbrace 1,2 \rbrace)=r(\lbrace 1,3 \rbrace)=r(\lbrace 1,4 \rbrace)=r(\lbrace 2,3 \rbrace)=r(\lbrace 2,4 \rbrace)=3$ and $r(\lbrace 3,4 \rbrace)=r(\lbrace 1,2,3 \rbrace)=r(\lbrace 1,2,4 \rbrace)=r(\lbrace 1,3,4 \rbrace)=r(\lbrace 2,3,4 \rbrace)=r(\lbrace 1,2,3,4 \rbrace)=4$. The rank function $r$ does not satisfy the Ingleton inequality \cite{ingleton1971representation} which is a necessary condition for discrete polymatroid to be representable. Hence this discrete polymatroid is not representable. The set of basis vectors for this discrete polymatroid is given by 
\begin{flalign*}
\lbrace (0,0,2,2),(2,1,1,0),(2,1,0,1),(2,0,1,1),(0,2,1,1),& \\ (1,2,0,1),(1,2,1,0),(1,1,2,0),(1,0,2,1),(1,1,0,2),\\(1,0,1,2),(0,1,1,2),(1,1,1,1) \rbrace.
\end{flalign*}
\end{example}

\begin{lemma}
Consider a discrete polymatroid $\mathbb{D}$ on the ground set $\lceil m \rfloor$, with a representation $V_1,V_2,\dotso,V_m$. Each $V_{i}$ can be expressed as the column span of a $r(\lceil m \rfloor) \times r(i)$ matrix $A_{i}$. Let $A$ be the concatenated matrix $[A_{1}~A_{2} \ldots A_{r}]$. The following operations on $A$ does not change the discrete polymatroid $\mathbb{D}$ : (i) Interchange two rows, (ii) Multiply a row by a non-zero member of $\mathbb{F}_{q}$, (iii) Replace a row by the sum of that row and another, (iv) Delete a zero row (unless it is the only row), and (v)  Multiply a column by a non-zero member of $\mathbb{F}_{q}$.

\begin{proof}
Refer Appendix \ref{App:LemmaProof}.
\end{proof}
\end{lemma}

\begin{definition}
Consider a discrete polymatroid $\mathbb{D}=(\lceil m \rfloor),r)$. For $T \subseteq \lceil m \rfloor$, the \textit{contraction} of $T$ from $\mathbb{D}$ is $\mathbb{D}/T=(\lceil m \rfloor - T, r_{\mathbb{D}/T})$, with $r_{\mathbb{D}/T}(X)=r(X\cup T) - r(T)$.
\end{definition}

\begin{example}
\label{eg:DPMContraction1}
Let $\mathbb{D}$ be the discrete polymatroid in Example \ref{eg:DPM1}. The ground set of $\mathbb{D}=E(\mathbb{D})= \lceil 3 \rfloor$. The contraction of the set $T=\lbrace 3 \rbrace$ from $\mathbb{D}$ is the discrete polymatroid $\mathbb{D}/T=(\lbrace 1,2 \rbrace, r_{\mathbb{D}/T})$ where $r_{\mathbb{D}/T}$ is as follows: $r_{\mathbb{D}/T}(\lbrace 2 \rbrace)=1$ and $r_{\mathbb{D}/T}(\lbrace 1 \rbrace)=r_{\mathbb{D}/T}(\lbrace 1,2 \rbrace)=2$.
\end{example}

\begin{example}
\label{eg:DPMContraction2}
Consider the discrete polymatroid $\mathbb{D}$ of Example \ref{eg:DPM2}. Let $T_{1}=\lbrace 3 \rbrace$ and $T_{2}= \lbrace 4 \rbrace$ be the substets of ground set $E(\mathbb{D})$. The contraction of $T_{1}$ from $\mathbb{D}$ is the discrete polymatroid $\mathbb{D}/T_{1}=(\lbrace 1,2,4 \rbrace, r_{\mathbb{D}/T_{1}})$ where $r_{\mathbb{D}/T_{1}}$ is as follows: $r_{\mathbb{D}/T_{1}}(X)=1$ if  $X \in \lbrace \lbrace 1 \rbrace, \lbrace 2 \rbrace \rbrace$ and $r_{\mathbb{D}/T_{1}}(X)=2$ otherwise. The contraction of $T_{2}$ from $\mathbb{D}$ is the discrete polymatroid $\mathbb{D}/T_{2}=(\lbrace 1,2,3 \rbrace,r_{\mathbb{D}/T_{2}})$ where $r_{\mathbb{D}/T_{2}}(X)=0$ if $ X = \lbrace 1 \rbrace$ and $r_{\mathbb{D}/T_{2}}(X)=1$ otherwise.
\end{example}

\begin{lemma}
For disjoint subsets $T_{1}$ and $T_{2}$ of ground set of $\mathbb{D}, (\mathbb{D}/T_{1})/T_{2}=(\mathbb{D}/T_{2})/T_{1}=\mathbb{D}/(T_{1} \cup T_{2})$.
\begin{proof}
Refer Appendix \ref{App:LemmaProof}.
\end{proof}
\end{lemma}
Next we consider the contraction of an $\mathbb{F}$-representable discrete polymatroid.
\begin{lemma}
\label{lemma:discretePolymatroidContraction}
 Consider a discrete polymatroid $\mathbb{D}$ on ground set $E=\lceil m \rfloor$. Consider an element $e \in E$. There exists a representation  $V_{1},V_{2},\ldots,V_{m}$, such that the vector space $V_{e}$ corresponding to the representation of $e$ can be expressed as the column space of a $r(E) \times r(e)$ matrix $A_{e}$ which has only unit vectors in its columns. Let $A_{i}$ be the $r(E) \times r(i)$ matrix having $V_{i}$ as its column space. For all $i \in E-\lbrace e \rbrace$ obtain the matrix $A'_{i}$ from $A_{i}$ by deleting the rows corresponding to the non-zero entries in $A_{e}$. Let $V'_{i}$ be the column space of the matrix $A'_{i}$. The vector spaces $V'_{i}, i \in E-e$ forms the representation of the discrete polymatroid $\mathbb{D}/e$.

\begin{proof}
Refer Appendix \ref{App:LemmaProof}.
\end{proof}
\end{lemma}

\begin{example}
Consider the discrete polymatroid $\mathbb{D}$ of Example \ref{eg:DPM1}. In Example \ref{eg:DPMContraction1}, the rank function of the contracted matroid $\mathbb{D}/\lbrace 3 \rbrace$ is obtained. The representation of $\mathbb{D}$ is given in Example \ref{eg:RDPM1}. Note that the matrix $A_{3}= \left[\begin{array}{cc}
0 \\
0 \\
1 
\end{array}\right]$ has only unit vector in its column. By deleting the third column from $A_{1}$ and $A_{2}$ we obtain $A_{1}'= \left[\begin{array}{cc}
1 & 0\\
0 & 1
\end{array}\right]$ and $ A_{2}'=\left[\begin{array}{cc}
0 & 1\\
0 & 1
\end{array}\right]$. Let $V_{1}'$ and $V_{2}'$ be the column spaces of the matrix $A_{1}'$ and $A_{2}'$. It can be verified that the vector spaces $V_{1}'$ and $V_{2}'$ form the representation of the discrete polymatroid $\mathbb{D}/\lbrace 3 \rbrace$.
\end{example}

\begin{example}
Consider the discrete polymatroid $\mathbb{D}$ of Example \ref{eg:DPM2}. The contraction of $\lbrace 3 \rbrace$ and $ \lbrace 4 \rbrace$ from $\mathbb{D}$ is obtained in Example \ref{eg:DPMContraction2}. In the representation of $D$ given in Example \ref{eg:DPM2}, element $\lbrace 3 \rbrace$ is represented by unit vector. The representation for $\mathbb{D}/\lbrace 3 \rbrace$ is obtained by removing third row from the matrices representing other elements. The matrices obtained are $A_{1}'= \left[\begin{array}{cc}
1 \\
0 
\end{array}\right], A_{2}'=\left[\begin{array}{cc}
0\\
1
\end{array}\right]$ and $A_{4}'= \left[\begin{array}{cc}
1 & 0\\
0 & 1
\end{array}\right]$. It can be verified that the column spaces of the matrices above forms a representation of the matroid $\mathbb{D}/\lbrace 3 \rbrace$. The element $\lbrace 4 \rbrace$ of $E(\mathbb{D})$ is represented by $A_{4}=\left[\begin{array}{cc}
1 & 0\\
0 & 1\\
0 & 1
\end{array}\right]$. Note that the element is not represented by unit vectors alone. However by performing a row operation we can obtain a new representation in which the representation of $\lbrace 4 \rbrace$ is made up of unit vectors. An alternate representation for the discrete polymatroid $\mathbb{D}$ is given by the column spaces of the following matrices : $A_{1}= \left[\begin{array}{cc}
1 \\
0 \\
0 
\end{array}\right],A_{2}= \left[\begin{array}{cc}
0 \\
1 \\
1 
\end{array}\right],A_{3}= \left[\begin{array}{cc}
0 \\
0 \\
1 
\end{array}\right]$ and $A_{4}=\left[\begin{array}{cc}
1 & 0\\
0 & 1\\
0 & 0
\end{array}\right]$. The representation of the discrete polymatroid $\mathbb{D}/\lbrace 4 \rbrace$ can be obtained by removing the first two rows. Consider the matrices $A_{1}'=[0],A_{2}'=[1]$ and $A_{3}'=[1]$. The matrices $A_{1}',A_{2}'$ and $A_{3}'$ forms a representation of the discrete polymatroid $\mathbb{D}/\lbrace 4 \rbrace$.
\end{example}

\section{Error Correcting Index Codes and Discrete Polymatroids}
\label{Sec:ECICandDPM}

In this section, we establish a connection between vector linear differential error correcting index codes and representable discrete polymatroids. Consider a vector linear differential error correcting index code $\mathfrak{C}$ of length $c$ and dimension $n$ for an index coding problem $\mathcal{I}(X,\mathcal{R})$. The error correcting index code $\mathfrak{C}$ should be able to correct $\delta_{i}$ errors for receiver $R_{i}$. The discrete polymatroid which we obtain has a ground set $\lceil m+2c \rfloor$. Let the set $\lceil m+2c \rfloor - \lceil m+c \rfloor$ be denoted by $S(\mathfrak{C})$.

The following theorem gives a set of necessary and sufficient conditions for the existence of a vector linear differential error correcting index code of length $c$ and dimension $n$ for an index coding problem $\mathcal{I}(X,\mathcal{R})$.

\begin{theorem}
\label{thm:ECICandDPM}
A vector linear differential error correcting index code over $\mathbb{F}_{q}$ of length $c$ and dimension $n$ exists for an index coding problem $\mathcal{I}(X,\mathcal{R})$, iff there exists a discrete polymatroid $\mathbb{D}$ representable over $\mathbb{F}_{q}$ on the ground set $\lceil m+2c \rfloor$ with $rank(\mathbb{D})=mn+c$ satisfying the following conditions.
\begin{list}{•}{•}
\item (A) $r(\lbrace i \rbrace)=n, \forall i \in \lceil m \rfloor, r(\lceil m + c \rfloor)= mn+c , r(\lbrace m+i \rbrace)=1, \forall ~i \in \lceil 2c \rfloor$ .
\item \begin{flalign*} (B) r(\lceil m \rfloor \cup \lbrace m+i, m+c+i \rbrace) & = r(\lceil m \rfloor \cup \lbrace m+i \rbrace) \\ & = r(\lceil m \rfloor \cup \lbrace m+c+i \rbrace) \end{flalign*}
\item (C) For each receiver $R_{i}=(x_{f(i)},H_{i})$ and for each error pattern $\mathcal{F}=\lbrace e_{i_{1}},e_{i_{2}},\ldots,e_{i_{2\delta_{i}}}\rbrace$, let \[
T_{\overline{\mathcal{F}},i}= \lceil m+c \rfloor - \overline{H_{i}} - \lbrace m+i_{1},m+i_{2},\ldots,m+i_{2\delta_{i}} \rbrace. \] Let $\mathbb{D}_{\mathcal{F},i}$ be the $|\overline{H_{i}}| + c + 2\delta_{i}$ element matroid $\mathbb{D}/T_{\overline{\mathcal{F}},i}$. Then at every receiver $R_{i}=(x_{f(i)},H_{i})$ and for each valid error pattern $ \mathcal{F}$ we must have 
\begin{flalign*}
r_{\mathbb{D}_{\mathcal{F},i}}( \lbrace f(i) \rbrace \cup S(\mathfrak{C}))=r_{\mathbb{D}_{\mathcal{F},i}}(S(\mathfrak{C})).
\end{flalign*}
\end{list}

\begin{IEEEproof}
First we prove the ``only if'' part. Suppose there exists a vector linear differential error correcting index code $\mathfrak{C}$ of length $c$ over $\mathbb{F}_{q}$ for the index coding problem $\mathcal{I}(X,\mathcal{R})$. For $k \in \lceil m \rfloor,$ let $A_{k}$ be the $(mn+c) \times n$ matrix with the $(i,j)^{th}$ entry being one for $i=(k-1)n+t, j=t,$ where $t \in \lceil n \rfloor$ and all other entries being zero. For $i \in \lceil c \rfloor$, let $A_{m+i}$ be the vector $\epsilon_{mn+i,mn+c}$. Since the index code $\mathfrak{C}$ is linear it can be represented by a $mn \times c$ matrix $L$. Let $\zeta$ be the concatenated matrix $\left(\begin{array}{c}
L\\
I_{c}
\end{array}\right)$. Note that order of $\zeta$ is $(mn+c) \times c$. Let $A_{m+c+i}= \zeta^{(i)}$, for $i \in \lceil c \rfloor$. Define $V_{i}$ to be the column span of $A_{i}$. We can verify that the discrete polymatroid $\mathbb{D}(V_{1},V_{2},\ldots,V_{m+2c})$ satisfies the conditions of theorem. Condition (A) holds as the vector spaces $V_{1},V_{2},\ldots, V_{m+c}$ are linearly independent. Also the vector spaces $V_{m+i}$ for $i \in \lceil 2c \rfloor$ are one dimensional. 

The element $m+c+i$ of the discrete polymatroid is represented by $V_{m+c+i}=<\zeta^{i}>$. Note that $\zeta^{(i)}=\left(\begin{array}{c}
L^{(i)}\\
I_{c}^{(i)}
\end{array}\right)$. The vector $L^{(i)}$ corresponds to a transmission of the index code and hence should be a linear combination of messages. The vector space $V_{m+i}$ is the column span of the vector $A_{m+i}$ which has a $1$ in the $(mn+i)^{th}$ position. From this we can conclude that the vector $\zeta^{(i)}$ lies in the linear span of $V_{1},V_{2},\ldots,V_{m}$ and $V_{m+i}$. 

Consider a receiver $R_{i}=(x_{f(i)},H_{i})$ and an error pattern $\mathcal{F}_{k}=\lbrace e_{i_{1}},e_{i_{2}},\ldots,e_{i_{2\delta_{i}}}\rbrace$. Let $I(\mathcal{F}_{k})=\lbrace i_{1},i_{2},\ldots i_{2\delta_{i}} \rbrace$ be the set of indices corresponding to the error pattern and let the set $\lbrace n+i_{1},n+i_{2},\ldots n+i_{2\delta_{i}} \rbrace$ be denoted as $n+I(\mathcal{F}_{k})$. From the definition of $\mathbb{D}_{\mathcal{F}_{k},i}$ we note that it is a representable discrete polymatroid with $|\overline{H_{i}}| + c + 2\delta_{i}$ elements. The representation can be obtained by the method in Lemma \ref{lemma:discretePolymatroidContraction}. Note that the representation of $\lbrace f(i) \rbrace$ in the contracted discrete polymatroid $\mathbb{D}_{\mathcal{F}_{k},i} $ is the vector space spanned by the columns of $\mathcal{I}_{D(R_{i})}$. The representation of ${m+c+i}$ in the discrete polymatroid $\mathbb{D}_{\mathcal{F}_{k},i}$ is the space spanned by the column 
\[\mathcal{Z}_{i} = \zeta_{\overline{\hat{H_{i}}} \cup (n+ I(\mathcal{F}))}^{i} = \left(\begin{array}{c}
L_{\overline{\hat{H_{i}}}}^{i}\\
I_{I(\mathcal{F}_{j})}^{i}
\end{array}\right). \]
Since we have a vector linear differential error correcting index code, from Lemma \ref{lemma:errorcorrection}, we have \[
\mathcal{I}_{D(R_{i})}^{(k)}\subseteq \left\langle \left(\begin{array}{c}
L_{\overline{\hat{H_{i}}}}\\
\mathbb{I}_{supp(\epsilon)}
\end{array}\right)\right\rangle, ~ \forall ~ k \in \lceil n \rfloor.
\]
Note that the matrix  $\left(\begin{array}{c}
L_{\overline{\hat{H_{i}}}}\\
\mathbb{I}_{supp(\epsilon)}
\end{array}\right)$ is the concatenated matrix  $[\mathcal{Z}_{1} ~ \mathcal{Z}_{2} ~ \ldots ~ \mathcal{Z}_{c}]$.  So we have $\mathcal{I}_{D(R_{i})}$ lies in the linear span of $[\mathcal{Z}_{1} ~ \mathcal{Z}_{2} ~ \ldots ~ \mathcal{Z}_{c}]$ and Condition (C) holds for error pattern $\mathcal{F}_{k}$ and receiver $R_{i}$. Since the receiver and error pattern was chosen arbitrarily this completes the ``only if'' part of the proof.

Now we have to prove the if part. Let $\mathbb{D}$ be the $\mathbb{F}_{q}$ representable discrete polymatroid of rank $mn+c$ which satisfies conditions (A), (B) and (C). From (A), it follows that there exists vector subspaces $V_{i}, i \in \lceil m \rfloor$ which can be written as the column span of $(mn+c) \times n$ matrices $A_{i}$ over $F_{q}$, with $rank(A_{i})=n$. Also there exists vector subspaces $V_{m+i},i \in \lceil 2c \rfloor$ which can be written as the column span of non-zero $(mn+c) \times 1$ vector. Consider the concatenated matrix $A=[A_{1}~A_{2}~\ldots ~A_{m+2c}]$. Since $rank(\lceil m+c \rfloor)=mn+c$, the concatenated matrix $A$ can be written as $[I_{mn+c} ~ \zeta]$. First we prove that there exists an $mn \times c$ matrix $L$ such that $\zeta = 
\left(\begin{array}{c}
L\\
I_{c}
\end{array}\right).$  From Condition (B) we have that the column vector representing $m+c+i$ lies in the linear span of vector spaces representing $\lceil m \rfloor$ and $m+i$. The vector representing $m+c+i$ is $\zeta^{i}$. We have \[
\zeta^{i}= \underset{j \in \lceil m \rfloor}{\sum} A_{j}Y_{i,j} + d_{i}A_{m+i},
\] for some $Y_{i,j} \in \mathbb{F}_{q}^{n}$ and $d_{j}$ in $\mathbb{F}_{q}$. Condition (B) also ensures that $d_{i} \neq 0, \forall ~i \in \lceil c \rfloor$. This also ensures that $A_{m+c+i} \neq A_{m+c+j}$ for distinct $i,j \in \lceil c \rfloor$. Arranging all $\zeta^{i}$, we get \[
\zeta = \left(\begin{array}{c}
L_{mn\times c}\\
K_{c \times c}
\end{array}\right),
\] where each column of $L$, $L^{(i)}$ is the concatenated vector $[Y_{i,1} ~ Y_{i,2} ~ \ldots ~ Y_{i,m} ]'$ and $K$ is a diagonal matrix with $d_{i} , 1\leq i \leq c$ as its diagonal entries. The discrete polymatroid $\mathbb{D}$ does not change if some row or some column of its representation is multiplied by a non-zero element of $\mathbb{F}_{q}$. The matrix $A$ is now of the form $A=\left(\begin{array}{cc}
I_{mn+c} & \zeta \end{array}\right)$. Consider the matrix $A'$ obtained from $A$ by multiplying the rows $\lbrace mn+1,mn+2,\ldots,mn+c \rbrace$ by the elements $\lbrace d_{1}^{-1},d_{2}^{-1},\ldots,d_{c}^{-1} \rbrace$ respectively and then multiplying columns $\lbrace mn+1,mn+2,\ldots,mn+c \rbrace$ by $\lbrace d_{1},d_{2},\ldots,d_{c} \rbrace$ respectively.The matrix $A'$ is of the form $\left(\begin{array}{cc}

I_{mn+c} & \zeta' \end{array}\right)$ where $\zeta' = \left(\begin{array}{c}
L_{mn\times c}\\
I_{c}
\end{array}\right)$. The matrix $A'$ is a representation for the discrete polymatroid $\mathbb{D}$ proving our claim. In the last part of the proof we show that the matrix $L$ corresponds to a  linear differential error correcting index code. 
\\
Consider a receiver $R_{i}=(x_{f(i)},H_{i})$ and an arbitrary error pattern $\mathcal{F}_{k}$. The representation of the discrete polymatroid $\mathbb{D}_{\mathcal{F}_{k},i}$ can be obtained by using Lemma \ref{lemma:discretePolymatroidContraction}. Note that the representation of $\lbrace f(i) \rbrace$ in the contracted discrete polymatroid $\mathbb{D}_{\mathcal{F}_{k},i} $ is the vector space spanned by columns of $\mathcal{I}_{D(R_{i})}$. The representation of ${m+c+i}$ in the discrete polymatroid $\mathbb{D}_{\mathcal{F}_{k},i}$ is the space spanned by column $\mathcal{Z}_{i} = \left(\begin{array}{c}
L_{\overline{\hat{H_{i}}}}^{i}\\
I_{I(\mathcal{F}_{k})}^{i}
\end{array}\right).$ Consider the matrix $\mathcal{Z}$ obtained by concatenating the representations of the elements in $S(\mathfrak{C})$. The matrix $\mathcal{Z}$ is equal to $[\mathcal{Z}_{1} ~ \mathcal{Z}_{2} \ldots \mathcal{Z}_{c}] = \left(\begin{array}{c}
L_{\overline{\hat{H_{i}}}}\\
I_{I(\mathcal{F}_{k})}
\end{array}\right).$
From Condition (C) we have $r_{\mathbb{D}_{\mathcal{F},i}}( \lbrace f(i) \rbrace \cup S(\mathfrak{C}))=r_{\mathbb{D}_{\mathcal{F},i}}(S(\mathfrak{C})).$ Hence the columns of $\mathcal{I}_{\mathbb{D}(R_{i})}$ lies in the linear span of $\left(\begin{array}{c}
L_{\overline{\hat{H_{i}}}}\\
I_{I(\mathcal{F}_{k})}
\end{array}\right)$.
As the choice of receiver and the error pattern was arbitrary, using Lemma 1 it is seen that the index code given by the matrix $L$ is differential error correcting. This completes the proof of the theorem.
\end{IEEEproof}

\end{theorem}

Theorem \ref{thm:ECICandDPM} establishes a link between vector linear differential error correcting index codes and a representable discrete polymatroid satisfying certain properties. In Example \ref{eg:diffECIC} below, we consider an example of a differential error correcting index coding problem with vector linear solution and show the representable discrete polymatroid associated with it. 

\begin{example}
\label{eg:diffECIC}
Consider an index coding problem $\mathcal{I}(X,\mathcal{R})$, with $X=\lbrace x_{1},x_{2},x_{3} \rbrace, x_{i} \in \mathbb{F}_{2}^{2}$ and with $\mathcal{R}=\lbrace R_{1},R_{2},R_{3} \rbrace$. Let $R_{1}=(x_{1},H_{1}=\lbrace x_{2} \rbrace),R_{2}=(x_{2},H_{2}=\lbrace x_{1},x_{3} \rbrace)$ and $R_{3}=(x_{3},H_{3}=\lbrace x_{1},x_{2} \rbrace)$. Let $\delta_{1}=2$ and $\delta_{2}=\delta_{3}=1$. Consider the vector linear error correcting index code of length $13$ over $\mathbb{F}_{q}$ described by the $6 \times 13$ matrix 
\[
L=\left[\begin{array}{ccccccccccccccc}
1 & 0 & 1 & 0 & 1 & 0 & 1 & 0 & 0 & 1 & 0 & 0 & 0 \\
0 & 1 & 0 & 1 & 1 & 1 & 0 & 0 & 0 & 0 & 1 & 0 & 0 \\
0 & 0 & 0 & 0 & 1 & 0 & 0 & 1 & 0 & 1 & 0 & 0 & 0 \\
0 & 0 & 0 & 0 & 0 & 1 & 0 & 1 & 1 & 0 & 1 & 0 & 0 \\
0 & 0 & 1 & 0 & 0 & 0 & 1 & 1 & 0 & 0 & 0 & 1 & 0 \\
1 & 1 & 0 & 1 & 0 & 0 & 0 & 0 & 1 & 0 & 1 & 1 & 1 
\end{array}\right]
.\] Construct the concatenated matrix $\zeta=\left(\begin{array}{c}
L\\
I_{13}
\end{array}\right)$. Let $A_{l}, l \in \lceil 3 \rfloor$ denote the $6 \times 2$ matrix with the $(i,j)^{th}$ entry being one for $i=(l-1)2+t,j=t$ and all other entries being zeros. For $i \in \lceil 13 \rfloor$, let $A_{3+i}$ be the column vector of length $19$ with one in the $(6+i)^{th}$ entry and all other entries zero.  For $i \in \lceil 13 \rfloor$, let $A_{16+i}=\zeta^{(i)}$. Let $V_{i}, i \in \lceil 29 \rfloor$ denote the column span of $A_{i}$ over $\mathbb{F}_{2}$. The discrete polymatroid $\mathbb{D}(V_{1},V_{2},\ldots,V_{29})$ satisfies the conditions of Theorem \ref{thm:ECICandDPM}. 
\end{example}

\subsection{$\delta$-Error Correcting Index Codes}
Here we consider $\delta$-error correcting index codes in which all the receivers have the ability to correct $\delta$ number of errors. This is a special case of differential error correcting index code in which $\delta_{i}=\delta$ for all receivers $R_{i} \in \mathcal{R}$. We consider a single error correcting index coding problem and show the matroids associated with it in Example \ref{eg:singleerrorcorrecting} below.

\begin{example}
\label{eg:singleerrorcorrecting}
Consider an index coding problem $\mathcal{I}(X,\mathcal{R})$, with $X=\lbrace x_{1},x_{2},x_{3} \rbrace, x_{i} \in \mathbb{F}_{2}^{2}$ and with $\mathcal{R}=\lbrace R_{1},R_{2},R_{3} \rbrace$. Let $R_{1}=(x_{1},H_{1}=\lbrace x_{2},x_{3} \rbrace),R_{2}=(x_{2},H_{2}=\lbrace x_{1},x_{3} \rbrace)$ and $R_{3}=(x_{3},H_{3}=\lbrace x_{1},x_{2} \rbrace)$. Consider the vector linear error correcting index code of length $6$ over $\mathbb{F}_{q}$ described by the $6 \times 6$ matrix 
\[
L=\left[\begin{array}{ccccccccc}
1 & 0 & 0 & 1 & 1 & 1 \\
0 & 1 & 1 & 1 & 0 & 1 \\
1 & 0 & 1 & 0 & 1 & 0 \\
0 & 1 & 0 & 1 & 0 & 1 \\
1 & 0 & 1 & 0 & 1 & 0 \\
0 & 1 & 0 & 1 & 1 & 1
\end{array}\right]
.\] Construct the concatenated matrix $\zeta=\left(\begin{array}{c}
L\\
I_{6}
\end{array}\right)$. Let $A_{l}, l \in \lceil 3 \rfloor$ denote the $6 \times 2$ matrix with the $(i,j)^{th}$ entry being one for $i=(l-1)2+t,j=t$ and all other entries being zeros. For $i \in \lceil 6 \rfloor$, let $A_{3+i}$ be the column vector of length $12$ with one in the $(6+i)^{th}$ entry and all other entries zero.  For $i \in \lceil 6 \rfloor$, let $A_{9+i}=\zeta^{(i)}$. Let $V_{i}, i \in \lceil 15 \rfloor$ denote the column span of $A_{i}$ over $\mathbb{F}_{2}$. The discrete polymatroid $\mathbb{D}(V_{1},V_{2},\ldots,V_{15})$ satisfies the conditions of Theorem \ref{thm:ECICandDPM}. 
\end{example}

\subsection{Error Correction at only a subset of Receivers}

In this subsection we consider the case where only a specific subset of receivers require error correcting capability. Consider a subset $S$ of $\mathcal{R}$. Each receiver $R_{i} \in S$ should be able to correct $\delta_{i}$ errors. We can obtain error correcting only at particular subset $S$ of receivers, from a differential error correcting index code by setting $\delta_{i}=0, \forall R_{i} \notin S$. 

\section{Conclusion}
\label{Sec:Conclusion}
In this paper we consider a generalization of error correcting index codes in which each receivers have different error correcting capability. It was shown that vector linear differential error correcting index codes correspond to representable discrete polymatroid with certain properties. Our main theorem connects vector linear differential error correcting index codes to a representable discrete polymatroids. Using a non-representable discrete polymatroid satisfying the conditions of the theorem the possibility of non-linear error correcting codes could be explored.

\appendices
\section{Proofs of Lemmas in Section \ref{Sec:DiscretePolymatroid}}
\label{App:LemmaProof}
\textit{Lemma 2 :}
Consider a discrete polymatroid $\mathbb{D}$ on the ground set $\lceil m \rfloor$, with a representation $V_1,V_2,\dotso,V_m$. Each $V_{i}$ can be expressed as the column span of a $r(\lceil m \rfloor) \times r(i)$ matrix $A_{i}$. Let $A$ be the concatenated matrix $[A_{1}~A_{2} \ldots A_{r}]$. The following operations on $A$ does not change the discrete polymatroid $\mathbb{D}$.
\begin{itemize}
\item Interchange two rows.
\item Multiply a row by a non-zero member of $\mathbb{F}_{q}$.
\item Replace a row by the sum of that row and another.
\item Delete a zero row (unless it is the only row).
\item Multiply a column by a non-zero member of $\mathbb{F}_{q}$.
\end{itemize}
\begin{proof}
The columns of $A$ are the concatenated columns of $A_{i}$. Multiplying a column of $A_{i}$ by a non-zero member of $F_{q}$ does not change the vector space $V_{i}$ associated with it. Thus the discrete polymatroid remains same. We have $dim(\sum_{i \in X} V_i)=r(X),$ $\forall X \subseteq \lceil m \rfloor.$ The $dim(\sum_{i \in X} V_i)$ is equal to the rank of concatenated matrix $A_{X}=[A_{i}],i \in X$. The row operations specified above does not change the rank of $A_{X}, \forall X \subseteq \lceil m \rfloor$. Thus the dependencies among the new vector spaces will remain the same and hence the discrete polymatroid $\mathbb{D}$ remains same.
\end{proof}
\textit{Lemma 3 :}
For disjoint subsets $T_{1}$ and $T_{2}$ of ground set of $\mathbb{D}, (\mathbb{D}/T_{1})/T_{2}=(\mathbb{D}/T_{2})/T_{1}=\mathbb{D}/(T_{1} \cup T_{2})$.
\begin{proof}
We show that the discrete polymatroids have the same rank function. For $X \subseteq E-(T_{1} \cup T_{2})$, we have $r_{\mathbb{D}/(T_{1} \cup T_{2})}(X)= r(X \cup (T_{1} \cup T_{2})) - r(T_{1} \cup T_{2})$. For the discrete polymatroid $\mathbb{D}/T_{1},r_{\mathbb{D}/T_{1}}(X)=r(X \cup T_{1})-r(T_{1})$. Then 
\begin{flalign*}
r_{(\mathbb{D}/T_{1})/T_{2}}(X)& =r_{\mathbb{D}/T_{1}}(X \cup T_{2})-r_{\mathbb{D}/T_{1}}(T_{2}) &\\ &= r(X \cup T_{2} \cup T_{1}) - r(T_{2} \cup T_{1})\\ & = r_{\mathbb{D}/(T_{1} \cup T_{2})}(X).
\end{flalign*}
Similarly we can prove that the polymatroid $(\mathbb{D}/T_{2})/T_{1}$ has the same rank function.
\end{proof}
\textit{Lemma 4: }
 Consider a discrete polymatroid $\mathbb{D}$ on ground set $E=\lceil m \rfloor$. Consider an element $e \in E$. There exists a representation  $V_{1},V_{2},\ldots,V_{m}$, such that the vector space $V_{e}$ corresponding to the representation of $e$ can be expressed as the column space of a $r(E) \times r(e)$ matrix $A_{e}$ which has only unit vectors in its columns. Let $A_{i}$ be the $r(E) \times r(i)$ matrix having $V_{i}$ as its column space. For all $i \in E-\lbrace e \rbrace$ obtain the matrix $A'_{i}$ from $A_{i}$ by deleting the rows corresponding to the non-zero entries in $A_{e}$. Let $V'_{i}$ be the column space of the matrix $A'_{i}$. The vector spaces $V'_{i}, i \in E-e$ forms the representation of the discrete polymatroid $\mathbb{D}/e$.

\begin{proof}
Without loss of generality we can assume the element to be contracted to be $1$. Consider a set $X \subseteq E-\lbrace 1 \rbrace$. Construct the concatenated matrix $A_{X}=[A_{1} ~ A_{i}], i \in X$. The concatenated matrix $A_{X}$ will be of the form $\left[\begin{array}{cc}
I_{r(1)} & B\\
\mathcal{O} & A'_{X}
\end{array}\right]$ where $\mathcal{O}$ represents the zero matrix. From $A_{X}$ it is clear that  $rank(A'_{X})=rank(A_{X})- r(1)$. For the discrete polymatroid $\mathbb{D}/1$, $r_{D/1}(X) = r_{D}(X \cup \lbrace 1 \rbrace) - r_{D}(\lbrace 1 \rbrace)$. Since $X$ was chosen arbitrarily we can conclude that matrices $A'_{i}, i \in E-\lbrace 1 \rbrace$ is a representation of the discrete polymatroid $\mathbb{D}/1$.
\end{proof}

\end{document}